\documentclass[journal]{IEEEtran}

\usepackage{graphicx,amssymb,lineno}
\usepackage{amsmath,amsfonts,amssymb}

\usepackage{algorithmic}
\usepackage[usenames]{color}
\usepackage{float}

\usepackage[linesnumbered,ruled,vlined]{algorithm2e}

\newcommand*{\TitleFont}{%
	\usefont{\encodingdefault}{\rmdefault}{}{n}%
	\fontsize{20}{20}%
	\selectfont}

\usepackage{graphicx,graphics,color,epsfig,subfigure,graphpap,rotate}
\usepackage{times, verbatim, subfigure, epsfig, graphicx, latexsym, amsmath}
\usepackage{url}
\usepackage{subfigure}
\usepackage{booktabs}
\usepackage{CJKutf8}
\begin{document}
		\title{\TitleFont Robust Train-to-Train Transmission Scheduling in mmWave Band for High Speed Train Communication Systems}
		
		\author{Yunhan~Ma,
			Yong~Niu,~\IEEEmembership{Member,~IEEE},
			Shiwen~Mao,~\IEEEmembership{Fellow,~IEEE},
			Zhu~Han,~\IEEEmembership{Fellow,~IEEE},
			Ruisi~He,~\IEEEmembership{Member,~IEEE},
			Zhangdui~Zhong,~\IEEEmembership{Fellow,~IEEE},
			Ning~Wang,~\IEEEmembership{Member,~IEEE},
			and Bo~Ai,~\IEEEmembership{Fellow,~IEEE}

\thanks{Copyright (c) 2015 IEEE. Personal use of this material is permitted. However, permission to use this material for any other purposes must be obtained from the IEEE by sending a request to pubs-permissions@ieee.org. This work was supported by the National Key Research and Development Program of China under Grant 2021YFB2900301; by the National Key Research and Development Program of China under Grant 2020YFB1806903; in part by the Fundamental Research Funds for the Central Universities under Grant 2022JBXT001 and Grant 2022JBQY004; in part by the National Natural Science Foundation of China under Grant 62221001, Grant 62231009, Grant U21A20445; in part by the Fundamental Research Funds for the Central Universities 2023JBMC030; in part by NSF CNS-2107216, CNS-2128368, CMMI-2222810, ECCS-2302469, US Department of Transportation, Toyota and Amazon. (\emph{Corresponding authors: Y. Niu, B. Ai.})}

\thanks{Y. Ma, R. He, B. Ai, Z. Zhong are with the State Key Laboratory of Advanced Rail Autonomous Operation, Beijing Jiaotong University, Beijing 100044, China, and also with the Beijing Engineering Research Center of High-speed Railway Broadband Mobile Communications, Beijing Jiaotong University, Beijing 100044, China (e-mails: 21120110@bjtu.edu.cn, ruisi.he@bjtu.edu.cn, boai@bjtu.edu.cn, zhdzhong@bjtu.edu.cn).}

\thanks{Y. Niu is with the State Key Laboratory of Advanced Rail Autonomous Operation, Beijing Jiaotong University, Beijing 100044, China, and also with the National Mobile Communications Research Laboratory, Southeast University, Nanjing 211189, China (e-mail: niuy11@163.com).}

\thanks{S. Mao is with the Department of Electrical and Computer Engineering, Auburn University, Auburn, AL 36849-5201 USA (e-mail: smao@auburn.edu).}

\thanks{Z. Han is with the Department of Electrical and Computer Engineering at the University of Houston, Houston, TX 77004 USA, and also with the Department of Computer Science and Engineering, Kyung Hee University, Seoul, South Korea, 446-701 (e-mail: hanzhu22@gmail.com).}

\thanks{N. Wang is with the School of Information Engineering, Zhengzhou University, Zhengzhou 450001, China (e-mail: ienwang@zzu.edu.cn).}	
	}
		\maketitle
		\vspace{-1.5cm}
\begin{abstract}
Demands for data traffic in high-speed railway (HSR) has increased drastically. The increasing entertainment needs of passengers, safety control information exchanges of trains, etc., make train-to-train (T2T) communications face the challenge of achieving high-capacity and high-quality data transmissions. In order to greatly increase the communication capacity, it is urgent to introduce millimeter wave (mmWave) technology. Faced with the problem that mmWave link is easy to be blocked, this paper leverages the existing equipment to assist relay, and proposes an effective transmission scheduling scheme to improve the robustness of T2T communication systems. First of all, we formulate a mixed integer nonlinear programming (MINLP) optimization problem the transmission scheduling in T2T communication systems where mobile relays (MRs) are all working in the full-duplex (FD) mode. Then we propose a low complexity heuristic algorithm to solve the optimization problem, which consists of three components: relay selection, transmission mode selection, and transmission scheduling. The simulation results show that the proposed algorithm can greatly improve the number of completed flows and system throughput. Finally, we analyze the influence of different design parameters on the system performance. The results show that the proposed algorithm can achieve more data flows and system throughput within a reasonable communication distance threshold in T2T communication with obstacles in different orbits. It can balance the computational complexity and system performance to achieve an efficient and robust data transmission.
\end{abstract}
\section{Introduction}
		
Wireless communication for high-speed railway (HSR) is developing rapidly. As the basis of railway communication technology, the Global System for Mobile Communications–Railway (GSM-R) mainly carries voice services and a small amount of data services at a limited bit rate, which can hardly meet the requirements of intelligent railway services \cite{He}. Passenger broadband Internet access, entertainment applications, video sharing, train control and management systems, remote operation, monitoring and other bandwidth intensive applications are challenging the traditional railway communication system \cite{Karimi}. Therefore, it is urgent to develop 5G-railway (5G-R) to meet the diversified demands of railway wireless communications \cite{Sneps-Sneppe1}, \cite{Sneps-Sneppe2}.
		
The explosive growing demand for wireless services in the HSR, such as high-definition videos, online interactive games and streaming media business, has promoted the research on wireless communication in digital railway systems. At present, there are some existing work focusing on in-train communication links, aiming to support the on-board sensor network and the Internet access of passengers \cite{Zhang1}. It allows device-to-device (D2D) communications of users in the train. If train-to-train (T2T) communications for trains on different tracks is further enabled, more in-train users can be served with D2D communication. This can make better use of the direct transmission links between mobile users to ease the burden on the base station, realize the increasing interconnection of infrastructure, trains and passengers \cite{HORIZON}, and promote the development of the railway into a new era of ``intelligent mobile railway''. In addition, in the concept of modern train, it is necessary to not only improve the train-to-ground (T2G) communication capability, but also provide reliable T2T communications \cite{Unterhuber1}. New train operation related businesses and applications such as the railway Internet of Things, railway multimedia scheduling, and high-definition video security monitoring will continue to emerge. Moreover, with the achievement of unmanned driving, automatic cruise control is urgently required \cite{Zhao1}, \cite{Zhao2}. The train monitoring data and other reliability data are exchanged between trains under the self-organized network to facilitate enhanced train control and management, to greatly improve the efficiency and safety of railway transportation \cite{Unterhuber2}, \cite{Lehner} and promote the process of railway modernization in the future. In the future, T2G communication system will not be able to meet the needs of future train operation related businesses and the large number of users gathered in HSR. Therefore, it is necessary to provide a robust T2T communication system that can achieve high capacity and high quality.
		
It is a great challenge to provide high-quality communications in high-speed environments to achieve better user experience and more reliable links. The mobile relay (MR) has a strong processing capability, which can avoid penetration loss, and mitigate the multipath effect and Doppler effect \cite{Yaacoub}. Therefore, we consider deploying several MRs on the top of the train carriage. The MRs are used for the communication between trains to provide high-quality service and avoid train penetration loss \cite{Zhang1}.
		
Facing the T2T scenario with high data demand in the future, it is urgent to leverage the millimeter wave (mmWave) technology. mmWave communication has become a promising technology for 5G due to its rich spectrum resources. Its capacity is far higher than the existing microwave band wireless local area network (WLAN) and cellular mobile communications \cite{Xia}, \cite{Sun}. However, compared with the communication systems using lower carrier frequency, mmWave communication suffers huge propagation loss. The free space propagation loss is proportional to the square of the carrier frequency. For the wavelength of about 5mm, the free space propagation loss at 60GHz is 28dB more than that at 2.4GHz \cite{Singh1}. In general, the mmWave system will use highly oriented antennas to compensate for the large path loss at high frequencies \cite{Niu2}. Moreover, the wavelength of mmWave signals is very short, and the links in the 60GHz band is very sensitive to obstacles. For example, blockage by human bodies can reduce the link budget by 20–30dB \cite{Singh2}. Moreover, as the distance between two radio devices increases, the probability of allowing LOS link propagation decreases exponentially \cite{Bai}.
		
The future intelligent railway scene will face a large number of train control information from different tracks, and a large number of communication service demands of train passengers on different tracks. It is necessary to study a scenario where trains on different tracks communicate with each other using the mmWwave band to achieve effective information exchange. Research on T2T communication can relieve the load pressure of the base station (BS), and can be applied to the scenario without BS. It is an important communication mode,such as D2D, V2V. This paper considers that trains always keep moving in the same direction in the process of train-to-train communication. There will be uncertain obstacles between trains, which will easily block mmWave links. It is proposed to use suitable MRs to relay the blocked links. The cooperative communication with distributed MRs will significantly reduce the link outage probability \cite{Munjal}. Thus, reliable and high-quality communication links are provided to ensure the robust transmission of T2T communication systems. We also integrate full-duplex (FD) communication into the system, to allow the equipment to transmit and receive at the same time, and to provide a more efficient and flexible access strategy for multiple access \cite{Sabharwal}, \cite{Zhao3}. In this way, the transmission rate of MR can be doubled and the system capacity can be greatly improved \cite{Bhar}. The contributions of this paper are as follows:

\begin{itemize}		
\item We combine T2T communication with mmWave and FD mode. In the existing environment, MRs are used to overcome link blockage. The T2T communication system with high capacity and quality will be realized, with significantly improved efficiency.		

\item The optimization problem of T2T robust transmission scheduling is formulated. We also designed a low complexity heuristic algorithm to find a suitable relay for a blocked link in the T2T communication, which maximizes the number of flows to be transmitted, and realizes the robust transmission of the system.
		
\item Compared with the three existing baseline schemes, the proposed algorithm achieves a better performance. We also analyze the influence of different system parameters on the system performance.
\end{itemize}

The rest of this paper is arranged as follows. We first discuss the relevant work in Section~\ref{S2}. In Section~\ref{S3}, a T2T system model with obstacle blockage is established, and an example is given to illustrate its transmission scheduling scheme in a short time. Then in Section~\ref{S4}, we formulate an optimization problem for T2T communications. In Section~\ref{S5}, a heuristic transmission scheduling algorithm with low complexity is proposed to optimize the system performance. In Section~\ref{S6}, the proposed algorithm is compared with the three baseline schemes, and the performance under different system parameters is analyzed. Finally, we summarize this paper in Section~\ref{S7}.		
\section{Related Work}\label{S2}
		
Wireless communication in the field of HSR is developing rapidly, and operators are inclined to study the wireless standardized communication system of T2T and T2G communication \cite{Soliman1}. The European standard for T2T communications is ITS-G5. Unterhuber $ et\ al. $ \cite{Unterhuber2} equipped a train with ITS-G5 Cohda device and additional measuring equipment, analyzed the T2T performance of ITS-G5 in the railway environment, and investigated various scenarios from urban, suburban and rural areas. Garcia $ et\ al. $ \cite{Garcia} proposed a comprehensive channel model for the 400MHz inter-train communication system, including path loss, Doppler shift, fading, and delay spread, which provides a theoretical basis for the research of T2T communication networks.
		
Several existing works proposed to use mmWave for T2T communications. Soliman $ et\ al. $ \cite{Soliman2} conducted the world's first dynamic millimeter wave T2T propagation measurement study. The measurement results show that the multipath component reflected from the ground will change the received signal power, which can be effectively suppressed by the circularly polarized antenna. Zhao $ et\ al. $ \cite{Zhao1} used the mmWave band to achieve T2T communication, and studied the alignment of narrow beams in the workshop under the turning scene. The train position information exchanged by T2T communications is used to estimate the beam deflection angle and compared with the quantized phase angle to obtain accurate beam forming and combination vector. Zhao $ et\ al. $ \cite{Zhao2} proposed a MADRL based distributed resource allocation scheme for T2T communications, which can effectively reduce the interference and improve the throughput of T2T systems.
		
In the process of train-to-train communication, there will be uncertain obstacles between trains, and the millimeter wave link will be easily blocked. Some prior works indicate that devices can be used as relay for blocked links. Niu $ et\ al. $ \cite{Niu1} designed an efficient directed media access control protocol, where relay selection and spatial reuse were jointly optimized to overcome the blockage problem and improve network performance. The protocol performs better in latency and throughput, and achieves good fairness. Ma $ et\ al. $ \cite{Ma} proposed a  heuristic algorithm based on graph theory for the mmWave HSR communication system. UAV and MRs are used for relay assistance to effectively solve the blockage problem, schedule the flows that meet the QoS requirements and channel quality, and improve the number of completed flows and system throughput. Gao $ et\ al. $ \cite{Gao} designed a dual-hop relay aided emergency communication scheme based on UAV to solve the problem of high bit error rate (BER) in T2T systems.
		
While ensuring robustness, we also hope to expand the capacity of T2T communication systems. Zhang $ et\ al. $ \cite{Zhang2} proposed an SI elimination precoding algorithm based on OMP to eliminate residual FD mmWave SI and achieve high frequency spectral efficiency. Its spectral efficiency is about twice that of half-duplex (HD). At present, FD has been applied in high-speed railway systems. Zhang $ et\ al. $ \cite{Zhang3} proposed an SQP algorithm based on Lagrangian function to solve the bandwidth allocation problem between BS and FD MRs in the T2G communication system. Li $ et\ al. $ \cite{Li} proposed a joint resource allocation and computation offloading scheme using the FD MRs to solve the problems of user association, resource allocation and computing offload in the uplink train-to-ground communication scenario.
		
To sum up, there is no research on transmission scheduling for T2T mmWave communication on different tracks. In this paper, an efficient, robust transmission scheduling scheme is proposed for T2T communication systems.
		
\section{System Overview}\label{S3}
\subsection{System Model}
\begin{figure}[t]
	\begin{minipage}[t]{1\linewidth}
		\centering
		\includegraphics[width=1\columnwidth]{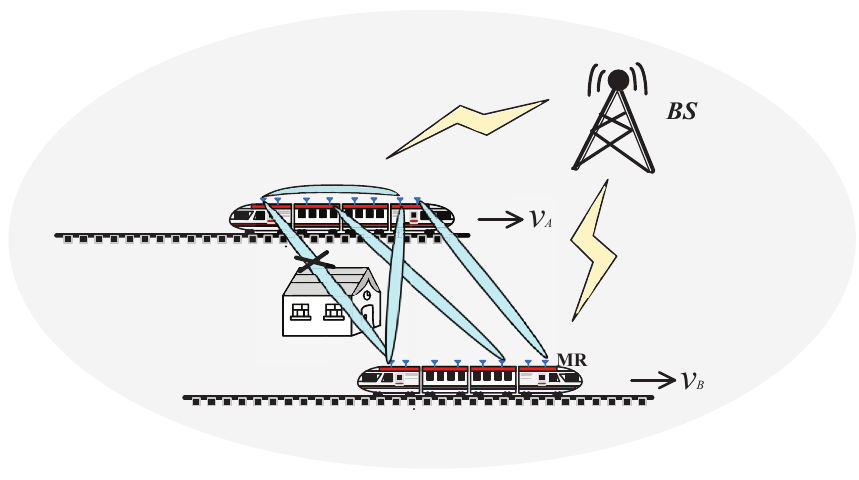}
	\end{minipage}%
		\caption{Millimeter-wave train-to-train communication system model}
		\label{fig:T2T} 
		\vspace*{-3mm}
\end{figure}

We consider a scenario of millimeter wave communication between two trains which can relieve the load of BS, as shown in Fig. \ref{fig:T2T}. BS is fixed outside the railway track. $ N $ mobile relays are evenly placed on top of each train to facilitate data exchange between trains. As the operation environment along the railway track is complex and changeable, buildings or terrain structures will cause random short-term occlusion to the wireless link. Therefore, with the continuous movement of the train, the channel conditions are also changing dynamically. There will be uncertain obstacles between them. Train control operation and the demand of a large number of users gathered in high-speed trains make the data demand between trains huge. For train-to-train communication, we only consider data flows sent to the other train. Each data flow has different traffic to be transmitted. In Fig. \ref{fig:T2T}, $ F $ flows ($ F\le 2{{N}^{2}} $) between trains need to be transmitted in the 28GHz frequency band. All MRs operate in the FD mode. In this paper, BS and all MRs are equipped with steerable directional antennas, allowing them to aim at relevant users to obtain higher antenna gain. In view of the problem of Doppler frequency shift caused by high-speed train movement and high carrier frequency of mmWave, the method in \cite{Ma} can be used to predict the powerful processing capacity of MR.
		
\begin{figure}[t]
	\begin{minipage}[t]{1\linewidth}
		\centering			\includegraphics[width=1\columnwidth]{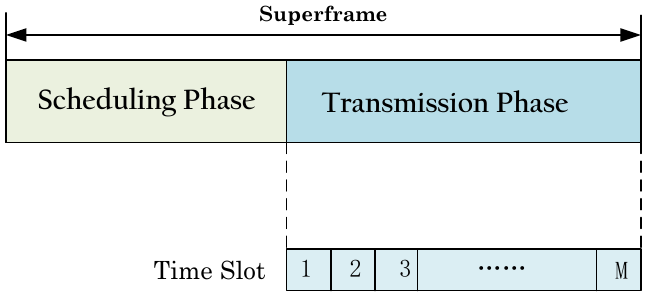}
	\end{minipage}%
	\caption{MAC transmission frame structure}
	\label{fig:superframe} 
	\vspace*{-3mm}
\end{figure}

The two trains in relative motion will not keep communicating with each other all the time. We need to set a communication distance threshold. When the distance between two trains exceeds this threshold, the poor channel conditions lead to the inability to communicate with each other. Thus, the time between the two trains moving from the initial position to the threshold distance is called the communication time of the trains. We will complete the scheduling of all flows transmission traffic within the communication time in combination with the link blockage. We propose to adopt a MAC transmission frame structure based on time division multiple access (TDMA) in the T2T communication system \cite{Dutta}. Then we divide the conmmunication time into a series of non-overlapping superframes, and each superframe consists of scheduling phase and transmission phase, as shown in Fig. \ref{fig:superframe}. In the scheduling phase of each frame, when the train-to-train communication link is blocked due to obstacles, MRs will notify the BS of the blocked information. Then BS will use appropriate MRs to assist link transmissions according to the blocked information and the traffic demands to be transmitted of all flows. Finally, all links are transmitted in the transmission phase of the frame.
		
In the transmission process of this paper, no matter the direct transmission mode or relay transmission mode, the transmitting node and receiving node of all links involved are MR. Therefore, there is only one type of link, that is, from MR to MR. The NLOS link due to obstacles will cause huge link loss, data loss and reception delay. Therefore, appropriate relays will be utilized to improve links quality, so that the NLOS links can be transformed into LOS links to meet the requirements of data transmissions. All links adopt the mmWave LOS path loss model \cite{Zhu}, \cite{Zhang3}. Since the channel quality between two high-speed trains will change rapidly during the relative driving process, we need to measure the link conditions during each time slot. Then the received power $ P_{r,l\left( i,j \right)}^{k} $ of the link $ l\left( i,j \right) $ in the $ k $-th timeslot can be expressed as
\begin{equation}
	P_{r,l\left( i,j \right)}^{k}={{k}_{0}}{{P}_{t}}G_{t\left( i,j \right)}^{k}G_{r\left( i,j \right)}^{k}d_{{{\left( i,j \right)}_{k}}}^{-n},
\end{equation}
where ${k}_{0}$ is a constant proportional to ${{\left( \frac{\lambda }{4\pi } \right)}^{2}}$, $\lambda$ is the wavelength of transmission signal, $ l\left( i,j \right) $ represents the link from transmit node $ i $ to receive node $ j $, $ G_{t\left( i,j \right)}^{k} $ and $ G_{r\left( i,j \right)}^{k} $ are the antenna gains of the transmitter and receiver, respectively, $n$ is the path loss index. $d_{{{\left( i,j \right)}_{k}}}^{-n}$ is the distance between node $ i $ and node $ j $ in the $ k $-th timeslot, and its value is dynamically changing and can be expressed as
\begin{equation}
	{{d}_{{{\left( i,j \right)}_{k}}}}=\left| \left( {{L}_{i}}-{{L}_{j}} \right)+\left( {{v}_{i}}-{{v}_{j}} \right)*k \right|,
\end{equation}
where $ L_{i} $ and $ L_{j} $ are the initial positions of node $ i $ and $ j $, respectively. $ v_{i} $ and $ v_{j} $ are the running speeds of the trains where node $ i $ and $ j $ are located.

Considering the interference of full-duplex and other parallel transmission links, the transmission rate $ R_{l\left( i,j \right)}^{k} $ of link $ l\left( i,j \right) $ in the $ k $-th timeslot can be expressed as
\begin{equation}
	R_{l\left( i,j \right)}^{k}=\eta W{{\log }_{2}}\left( 1+\frac{P_{r,l\left( i,j \right)}^{k}}{{{N}_{0}}W+\sum\limits_{h}{{{I}_{s}}+I_{l\left( i,j \right)}^{k}}} \right),\label{rate}
\end{equation}
where $\eta \in \left( 0,1 \right)$ is the efficiency of transceiver design, ${{N}_{0}}$ is the one-sided noise power spectral density of the Gaussian channel, and $ W $ is the channel bandwidth. When MR works in the FD mode, the self-interference (SI), which means that the signal sent by the transmitter will be received by the receiver of the same node \cite{Ahmed}. When the transmit power is $ P_{t} $, the FD self-interference $ {I}_{s} $ can be expressed as
		
\begin{equation}
	{{I}_{s}}=\beta {{P}_{t}},
\end{equation}
where $ \beta $ is SI cancellation parameter.

Then the second term in the denominator of (\ref{rate}) represents the self-interference effect on the FD device $ j $, where $ h $ represents the number of flows using node $ j $ as the transmitter in the same time slot. Since all MRs are single antenna devices, the value of $ h $ is either 0 or 1. The third item in the denominator indicates that in the $ k $-th timeslot, link $ l\left( i,j \right) $ will be interfered by parallel transmission from other links in the same timeslot. Since these links do not share any nodes with link $ l\left( i,j \right) $, the interference power that $ l\left( i,j \right) $ receives in the timeslot can be expressed as
\begin{equation}
	I_{l\left( i,j \right)}^{k}=\sum\limits_{p,q\ne i,j}{{{k}_{0}}{{P}_{t}}G_{t\left( p,j \right)}^{k}G_{r\left( p,j \right)}^{k}d_{{{\left( p,j \right)}_{k}}}^{-n}S_{l\left( p,q \right)}^{k}},
\end{equation}
where $ S_{l\left( p,q \right)}^{k} $ is a flag indicating whether link $ l\left( p,q \right) $ is scheduled in the current slot. If it is scheduled, then $ S_{l\left( p,q \right)}^{k}=1 $, which indicates that it interferes with link $ l\left( i,j \right) $. Otherwise $ S_{l\left( p,q \right)}^{k}=0 $.
		
In the T2T communication system, we divide all flows into relay assisted transmitting flows and direct transmitting flows, which are marked with subscripts $ a $ and $ b $, respectively. For a direct transmitting flow, the transmission rate of flow $ f\left( i,j \right) $ in the $ k $-th slot is equal to the rate of link $ l\left( i,j \right) $, which can be expressed as
\begin{equation}
	R_{b,f\left( i,j \right)}^{k}=R_{l\left( i,j \right)}^{k}.
\end{equation}
For the flow assisted by a relay, if flow $ f\left( i,j \right) $ uses node $ v_{ij} $ as relay, its transmission rate in the $ k $-th timeslot depends on the slower one of link $ l\left( i,v_{ij} \right) $ and link $ l\left( v_{ij},j \right) $, which can be expressed as
\begin{equation}
	R_{a{{v}_{ij}},f\left( i,j \right)}^{k}=\min \left\{ R_{l\left( i,{{v}_{ij}} \right)}^{k},R_{l\left( {{v}_{ij}},j \right)}^{k} \right\}.
\end{equation}
		
\subsection{An Example}
		
\begin{figure*}[t]
	\centering
	\includegraphics[width=2\columnwidth]{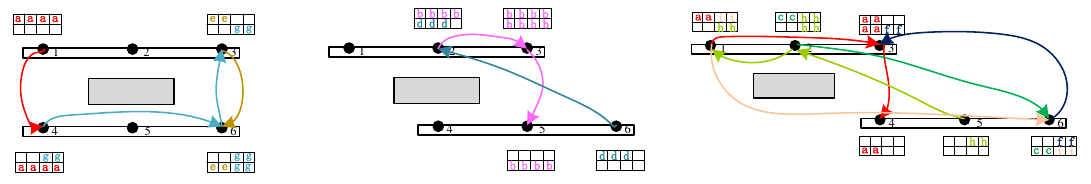}
	\caption{Example of two trains running in the same direction and conducting transmission scheduling within the communication time. It costs 3 MAC frames, and the scheduling schemes of the first, second, and third frame are shown from left to right in this figure. We set a grid for each node. The first row of the grid represents the transmission status of the node, and the second row represents its receiving status. a,b,c,d,e,f,g,h,i correspond to the data to be transmitted of all flows with traffic in the demand matrix D}
	\label{fig:example} 
	\vspace*{-3mm}
\end{figure*}
Fig. \ref{fig:example} shows an example of two moving trains communicating with each other. The gray square area indicates an obstacle. It is assumed that roof of train A, there are nodes $ MR1 $, $ MR2 $, and $ MR3 $, and on the roof of train B, there are nodes $ MR4 $, $ MR5 $, and $ MR6 $.
In order to reduce the load of BS, we make MR deployed on two trains directly transmit data to each other, which is used to transmit some train control and a large number of user data demands. A matrix \textbf{D} is used to represent the traffic demands to be transmitted between the MRs.
\begin{equation}
	\textbf{D}=\left(
	\begin{array}{cccccc}
		0 & 0 & 0 & 18 & 0 & 2\\
		0 & 0 & 0 & 0 & 12 & 4\\
		0 & 0 & 0 & 0 & 0 & 6\\
		0 & 0 & 5 & 0 & 0 & 0\\
		6 & 0 & 0 & 0 & 0 & 0\\
		0 & 10 & 5 & 0 & 0 & 0\\
	\end{array}
	\right).
\end{equation}
		
In this example, we assume that train A and train B start from the same initial position and move in the same direction, but the speed of train B is faster than that of train A. They can maintain the communication time of three frames. Each frame contains only four transmission time slots, which is very short. For convenience, we consider that the channel conditions remain unchanged within a frame, but vary between frames, depending on obstacles and relative distances. The following matrices $ C_{1} $, $ C_{2} $, and $ C_{3} $ represent the link transmission rate of the first, second, and third frame of MR respectively.
\begin{equation}
	C_{1}=\left(
	\begin{array}{cccccc}
		0 & 4 & 4 & 3 & 0 & 0\\
		4 & 0 & 4 & 0 & 0 & 0\\
		4 & 4 & 0 & 0 & 0 & 3\\
		3 & 0 & 0 & 0 & 4 & 4\\
		0 & 0 & 0 & 4 & 0 & 4\\
		0 & 0 & 3 & 4 & 4 & 0\\
	\end{array}
	\right)
\end{equation}
\begin{equation}
	C_{2}=\left(
	\begin{array}{cccccc}
		0 & 4 & 4 & 0 & 0 & 0\\
		4 & 0 & 4 & 0 & 0 & 3\\
		4 & 4 & 0 & 0 & 3 & 3\\
		0 & 0 & 0 & 0 & 4 & 4\\
		0 & 0 & 3 & 4 & 0 & 4\\
		0 & 3 & 3 & 4 & 4 & 0\\
	\end{array}
	\right)
\end{equation}
\begin{equation}
	C_{3}=\left(
	\begin{array}{cccccc}
		0 & 4 & 4 & 0 & 0 & 1\\
		4 & 0 & 4 & 0 & 3 & 2\\
		4 & 4 & 0 & 3 & 3 & 3\\
		0 & 0 & 3 & 0 & 4 & 4\\
		0 & 3 & 3 & 4 & 0 & 4\\
		1 & 2 & 3 & 4 & 4 & 0\\
	\end{array}
	\right).
\end{equation}
		
Then, according to the traffic demand matrix \textbf {D} and the link transmission rate matrix $ C_{1} $, $ C_{2} $, and $ C_{3} $ in each frame, we will use the following scheme to complete the transmission of all flows.
		
In $ D $, $ D\left( 1,4 \right)=18 $ indicates that the traffic to be sent from $ MR_{1} $ to $ MR_{4} $ is 18. We find that link $ l\left( 1,4 \right) $ is not blocked only in the first frame, but can only complete part of the traffic by spending all the slots in this frame. So we need to use other frames for transmission. According to the rate matrix, any node in the second frame as the relay of link $ l\left( 1,4 \right) $ will be blocked. Therefore, two time slots should be used for transmission in the third frame with the help of node $ MR_{3} $. $ D\left( 2,5 \right)=12 $ indicates that the traffic to be sent from $ MR_{2} $ to $ MR_{5} $ is 12. We can use node $ MR_{3} $ to relay in the second frame, which needs to occupy all time slots. $ D\left( 2,6 \right)=4 $ and $ D\left( 6,2 \right)=10 $ exist in the traffic matrix, which indicates that $ MR_{2} $ and $ MR_{6} $ have traffic for each other. However, the link between node $ MR_{2} $ and $ MR_{6} $ is blocked in the first frame, so direct transmission can only be used in the second frame and the third frame. The scheme in the figure shows that three time slots are occupied in the second frame to complete the flow $ f\left( 6,2 \right) $, and two time slots are occupied in the third frame to complete flow $ f\left( 2,6 \right) $. $ D\left( 3,6 \right)=6 $ and $ D\left( 6,3 \right)=5 $ indicate that $ MR_{3} $ and $ MR_{6} $ have traffic for each other. The link is not blocked in these three frames, but $ MR_{3} $ has been occupied all slots in the second frame. Therefore, two time slots should be used in the first frame and the third frame, so that the two flows can complete the direct transmission.

$ D\left( 4,3 \right)=5 $ indicates that the traffic to be sent from $ MR_{4} $ to $ MR_{3} $ is 5. At this time, the flow is blocked in the first frame and the second frame, and the receiving end of $ MR_{3} $ has been occupied in the second frame and the third frame, so the flow can only be transmitted in the relay mode in the first frame. We decide to use two time slots to complete the flow with the help of $ MR_{6} $. $ D\left( 5,1 \right)=6 $ indicates that the traffic to be sent from $ MR_{5} $ to $ MR_{1} $ is 6. At this time, the receiving and transmitting status of the MR in the first frame can no longer be used for relay. In the second frame, $ MR_{3} $ is also fully occupied, and there is no other suitable relay node. Therefore, we can only consider using node $ MR_{2} $ in the third frame to complete the flow $ f\left( 5,1 \right) $ in two time slots. $ D\left( 1,6 \right)=2 $ indicates that the traffic to be sent from $ MR_{1} $ to $ MR_{6} $ is 2. At this time, the link is blocked in the first frame and the second frame. Therefore, we occupy two time slots in the third frame and use the direct link to complete the flow.
		
From the simplified example, we can see that it is complex to schedule all flows when the traffic demand and link transmission rates are known. So in the T2T communication system, facing the dynamic and unknown channel environment and long transmission time, how to carry out relay selection and transmission scheduling to maximize the system performance is a key problem.
\section{Problem Formulation}\label{S4}
		
In the T2T communication system, it is assumed that $ N $ MRs are uniformly deployed on the roof of each train, and there are F ($ F\le 2{{N}^{2}} $) data flows to be sent to each other. Each data flow has two transmission modes: direct transmission mode and relay node transmission mode. We assume that each flow can only be transmitted in a specific mode in each frame, but different transmission modes can be selected in different frames. We divide the commnication time into several frames, and each frame is divided into $ T_ {s} $ scheduling timeslots and $ K $ transmission timeslots.
		
In the initial state, each flow has its own traffic demand $ q_{f\left( i,j \right)} $ to be transmitted. In the $ k $-th timeslot, we need to set two binary variables $ S_{a{{v}_{ij}},f\left( i,j \right)}^{k} $ and $ S_{b,f\left( i,j \right)}^{k} $ for each flow, which respectively indicate whether flow $ f\left( i,j \right) $ actually uses the relay assisted mode and direct mode to transmit in this timeslot. Where $ v_{ij} $ represents the relay node used by flow $ f\left( i,j \right) $ in the relay assisted mode. We use $ C_{f\left( i,j \right)}^{\left\lfloor \frac{k}{{{T}_{s}}+K} \right\rfloor } $ to indicate whether flow $ f\left( i,j \right) $ is completely transmitted in the $ \left\lfloor \frac{k}{{{T}_{s}}+K} \right\rfloor $-th frame.  If it is completed, $ C_{f\left( i,j \right)}^{\left\lfloor \frac{k}{{{T}_{s}}+K} \right\rfloor } $ will be 1; otherwise, $ C_{f\left( i,j \right)}^{\left\lfloor \frac{k}{{{T}_{s}}+K} \right\rfloor } $ will be 0. Therefore, when flow $ f\left( i,j \right) $ is transmitted in any mode in the $ k $-th slot and its remaining traffic $ q_{f\left( i,j \right)}^{k} $ is less than or equal to 0 at the end of the $ k $-th slot, it means that this flow has just been transmitted in the $ \left\lfloor \frac{k}{{{T}_{s}}+K} \right\rfloor $-th frame corresponding to slot $ k $. So we set $ C_{f\left( i,j \right)}^{\left\lfloor \frac{k}{{{T}_{s}}+K} \right\rfloor } $ to 1; When the above conditions are not met, the value is 0. This shown in constraint (\ref{CONS1}).
\begin{equation}
	\begin{aligned}
		&C_{f\left( i,j \right)}^{\left\lfloor \frac{k}{{{T}_{s}}+K} \right\rfloor }\\
		&=\begin{cases}
			1,&\mbox{if $ q_{f\left( i,j \right)}^{k}\le 0 $ and $ S_{a{{v}_{ij}},f\left( i,j \right)}^{k}+S_{b,f\left( i,j \right)}^{k}=1 $};\\
			0,&\mbox{otherwise}.
		\end{cases}
	\end{aligned}\label{CONS1}
\end{equation}
		
Constraint (\ref{CONS2}) represents the calculation process of the remaining traffic of flow $ f\left( i,j \right) $ at the end of the $ k $-th slot, which is equal to the remaining traffic at the end of the ($ k-1 $)-th timeslot minus the traffic transmitted in the $ k $-th timeslot. $ R_{a{{v}_{ij}},f\left( i,j \right)}^{k} $ and $ R_{b,f\left( i,j \right)}^{k} $  represent the transmission rates of the flow using node $ v_{ij} $ for relay and the direct mode, respectively. In addition, when $ k=1 $, $ q_{f\left( i,j \right)}^{0} $ in constraint (\ref{CONS2}) is equal to the initial traffic demand $ q_{f\left( i,j \right)} $.
\begin{equation}
	\begin{aligned}
		q_{f\left( i,j \right)}^{k}=&q_{f\left( i,j \right)}^{k-1}-S_{a{{v}_{ij}},f\left( i,j \right)}^{k}R_{a{{v}_{ij}},f\left( i,j \right)}^{k}\Delta t\\
		&-S_{b,f\left( i,j \right)}^{k}R_{b,f\left( i,j \right)}^{k}\Delta t.\\
	\end{aligned}\label{CONS2}
\end{equation}
		
In constraint (\ref{CONS3}), two binary variables $ a_{f\left( i,{{v}_{ij}},j \right)}^{t} $ and $ b_{f\left( i,j \right)}^{t} $ are defined for each flow, indicating that flow $ f\left( i,j \right) $ intends to use node $ v_{ij} $ to relay assisted mode or direct mode to transmit in the $ t $-th frame, respectively. The values of these four variables are either 0 or 1.
\begin{equation}
	a_{f\left( i,{{v}_{ij}},j \right)}^{t}, b_{f\left( i,j \right)}^{t}, S_{a{{v}_{ij}},f\left( i,j \right)}^{k}, S_{b,f\left( i,j \right)}^{k}\in \left\{ 0, 1 \right\}.\label{CONS3}
\end{equation}
		
Since each flow can only select one transmission mode in each frame, it is necessary to constrain the sum of $ a_{f\left( i,{{v}_{ij}},j \right)}^{t} $ and $ b_{f\left( i,j \right)}^{t} $ to be 1, as shown in constraint (\ref{CONS4}).
\begin{equation}
	a_{f\left( i,{{v}_{ij}},j \right)}^{t}+b_{f\left( i,j \right)}^{t}=1.\label{CONS4}
\end{equation}
		
In constraint (\ref{CONS5}), \textbf{P(·)} is an assignment operation, that is, all elements in the bracket must be equal to the value on the right of the equation. For any node $ i $, $ i^{t} $ and $ i^{r} $ represent the transmitting and receiving states of node $ i $, respectively. If the value is 1, it indicates that node $ i $ is being occupied at this time; The value of 0 indicates node $ i $ it is idle. Therefore, the necessary condition for any flow to be successfully scheduled is that both the transmitting status of its transmitter and the receiving status of its receiver are idle. If flow $ f\left( i,j \right) $ is expected to use node $ v_{ij} $ for relay transmission in the $ t $-th frame, and at the same time, the corresponding transmitting and receiving states of the relevant nodes in the current timeslot are idle, and there is still remaining traffic after the transmission of the previous slot is completed, the flow can choose whether to use node $ v_{ij} $ for relay transmission in the current slot. If the flow is to be transmitted using the direct link in frame $ t $, the corresponding node states are 0 in the current slot, and there is still residual traffic to be transmitted, then the flow can choose whether to conduct direct transmission in slot $ k $. Otherwise, the flow cannot be transmitted.
\begin{equation}
	\begin{aligned}
		&S_{a{{v}_{ij}},f\left( i,j \right)}^{k}\in \left\{ 0,1 \right\},\\
		&\;\;\;\;\mbox{if $ a_{f\left( i,{{v}_{ij}},j \right)}^{\left\lfloor \frac{k}{{{T}_{s}}+K} \right\rfloor }=1 $ and $ P\left( {{i}^{t}},v_{ij}^{t},v_{ij}^{t},{{j}^{r}} \right)=0 $ and $ q_{f\left( i,j \right)}^{k-1}>0 $};\\
		&S_{b,f\left( i,j \right)}^{k}\in \left\{ 0,1 \right\},\\
		&\;\;\;\;\mbox{if $ b_{f\left( i,j \right)}^{\left\lfloor \frac{k}{{{T}_{s}}+K} \right\rfloor }=1 $ and $ P\left( {{i}^{t}},{{j}^{r}} \right)=0 $ and $ q_{f\left( i,j \right)}^{k-1}>0 $};\\
		&S_{a{{v}_{ij}},f\left( i,j \right)}^{k}+S_{b,f\left( i,j \right)}^{k}=0,\;\;\;\;\mbox{otherwise}.\\
	\end{aligned}\label{CONS5}
\end{equation}

If it is in the scheduling phase of the frame, all flows cannot be transmitted in any way, and the transmitting and receiving states of all nodes are set to 0, as shown in constraint (\ref{CONS6}),
\begin{equation}
	\begin{aligned}
		&S_{a{{v}_{ij}},f\left( i,j \right)}^{k}+S_{b,f\left( i,j \right)}^{k}=0,P\left( {{i}^{t}},v_{ij}^{t},v_{ij}^{t},{{j}^{r}} \right)=0,\\
		&\;\;\;\mbox{if $ k=\left\lceil \frac{k}{{{T}_{s}}+K} \right\rceil \left( {{T}_{s}}+K \right)+u,u=1,2,\cdots ,{{T}_{s}} $}.\\
	\end{aligned}\label{CONS6}
\end{equation}
		
Constraint (\ref{CONS7}) indicates that if flow $ f\left ( i,j \right) $ is transmitted in the relay or direct mode in the previous timeslot, and the remaining traffic is less than or equal to 0 at the end of this slot, the flow has just completed its transmission. It should not be scheduled in the next timeslot, and the states of nodes involved in the flow should be released.
\begin{equation}
	\begin{aligned}
		&S_{a{{v}_{ij}},f\left( i,j \right)}^{k}=0,P\left( {{i}^{t}},v_{ij}^{t},v_{ij}^{t},{{j}^{r}} \right)=0,\\
		&\;\;\;\;\;\;\mbox{if $ q_{f\left( i,j \right)}^{k-1}\le 0 $ and $ S_{a{{v}_{ij}},f\left( i,j \right)}^{k-1}=1 $};\\
		&S_{b,f\left( i,j \right)}^{k}=0,P\left( {{i}^{t}},{{j}^{r}} \right)=0,\\
		&\;\;\;\;\;\;\mbox{if $ q_{f\left( i,j \right)}^{k-1}\le 0 $ and $ S_{b,f\left( i,j \right)}^{k-1}=1 $}.\\
	\end{aligned}\label{CONS7}
\end{equation}
		
If flow $ f\left( i,j \right) $ is transmitted in the relay or direct mode at this time slot, the corresponding node's transmitting and receiving states should be set to 1. For a flow relayed by node $ v_{ij} $, the transmitting state of node $ i $, the transmitting and receiving state of $ v_{ij} $, and the receiving state of $ j $ should be set to 1; For flow $ f\left( i,j \right) $ using direct transmission, the transmitting state of node $ i $ and the receiving state of node $ j $ should be set to 1, as shown in constraint (\ref{CONS8}),
\begin{equation}
	\begin{aligned}
		P\left( {{i}^{t}},v_{ij}^{t},v_{ij}^{t},{{j}^{r}} \right)=1,\;\;&\mbox{if $ S_{a{{v}_{ij}},f\left( i,j \right)}^{k}=1 $};\\
		P\left( {{i}^{t}},{{j}^{r}} \right)=1,\;\;&\mbox{if $ S_{b,f\left( i,j \right)}^{k}=1 $}.\\		
	\end{aligned}\label{CONS8}
\end{equation}

In this T2T transmission system, we need to set a threshold $ dis $ to indicate that two trains cannot communicate with each other. When the distance between two trains exceeds the threshold $ dis $, the communication between trains cannot be achieved. Then the number of timeslots for two trains to maintain communication can be calculated as follows,
\begin{equation}
	M=\left\lceil \frac{dis-{{L}_{t}}\pm\left( {{L}_{A}}-{{L}_{B}} \right)}{\left| {{v}_{A}}-{{v}_{B}} \right|*\Delta t} \right\rceil,\label{CONS9}
\end{equation}
where $ L_{t} $ represents the length of the train. $ L_{A} $ and $ v_{A} $ represent the initial position and speed of the train running in front, $ L_{B} $ and $ v_{B} $ represent the position and speed of the train following. When $ {{v}_{A}}>{{v}_{B}} $, take the plus sign in constraint (\ref{CONS9}), otherwise take the minus sign.

To maximize the number of flows transmitted when the two trains are moving relative to each other, we express the optimization problem ({\rm{P1}}) as follows,
\begin{equation}\hspace{0.8cm}
	({\rm{P1}})\;\;\;\;\max \sum\limits_{t=1}^{T}{\sum\limits_{i,j}{C_{f\left( i,j \right)}^{t}}},\;\;\;T=\left\lceil \frac{M}{{{T}_{s}}+K} \right\rceil,
\end{equation}
	\hspace{2.8cm}s.t.
	\hspace{0.15cm}Constraints (\ref{CONS1})--(\ref{CONS9}).
	
It can be seen that this is a mixed integer nonlinear programming (MINLP) problem, which is NP-hard. It contains many variables and has high computational complexity, and will take a lot of time to solve with an optimization software. A low complexity solution is urgently needed. Next, we propose a heuristic relay selection algorithm, transmission mode selection algorithm and transmission scheduling algorithm to obtain the suboptimal system performance.
\section{T2T Communication Scheduling Algorithm}\label{S5}
This article aims to overcome dynamic link occlusion and allocate appropriate transmission modes and time slots for all flows within the frame where two trains can maintain communication, in order to complete the transmission of all data streams as much as possible, thus providing an efficient and robust T2T communication.

Firstly, we refer to Algorithm 2: transmission mode selection algorithm. In the scheduling phase of the frame, all currently unblocked flows are screened out based on the dynamic blockage situation of current links. For remaining data flows that must be blocked currently, we follow Algorithm 1: relay node selection algorithm, select the most suitable node for relay assistance. The relay node needs to ensure that each hop of the relay flow can maintain unblocked communication for a long time within the frame. Finally, we use Algorithm 3 to prioritize transmitting all direct transmission flows based on the transmission and reception states of FD nodes, and then transmit the remaining relay flows as much as possible in ascending order of the required number of time slots.

\subsection{Relay Selection Algorithm}
		
\begin{figure}[t]
	\begin{minipage}[t]{1\linewidth}
		\centering
		\includegraphics[width=1\columnwidth]{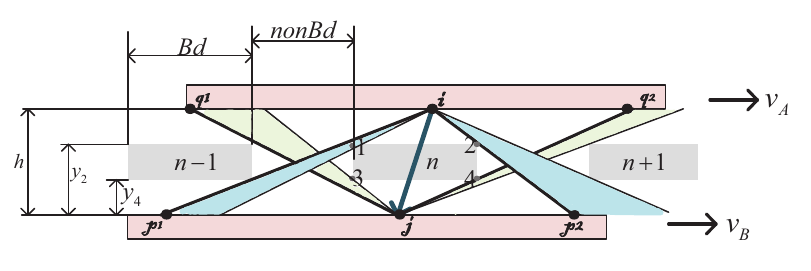}
	\end{minipage}%
	\caption{Simplified model diagram of the relay selection algorithm}
	\label{fig:simple} 
	\vspace*{-3mm}
\end{figure}
Fig. \ref{fig:simple} provides a simplified top view of the mmWave T2T communication model with certain obstacles. For convenience, we set a fixed-length obstacle at every fixed distance. There are several obstacles in the actual operation process, and we will use the ($ n-1 $)-th, $ n $-th, and ($ n+1 $)-th obstacles to explain the proposed algorithm, as shown in Fig. \ref{fig:simple}. The gray shadow part is covered by obstacles, and the length is expressed in $ Bd $; The blank part is an unobstructed area, and the length is represented by $ nonBd $. We combine the $ n $-th obstacle and the $ n $-th obstacle free area into the $ n $-th unit, whose total length is expressed by $ len $. Suppose that node $ i $ on train A is to send traffic to node $ j $ on train B, but this flow is blocked by the $ n $-th obstacle as Fig. \ref{fig:simple}. At this time, we should consider selecting an appropriate relay node to assist the transmission of this blocked flow. The following will describe how to find a suitable relay node to achieve the relatively good system performance.

\begin{algorithm}[t]
	\DontPrintSemicolon
	\caption{Relay selection algorithm}\label{alg:relay}
	\textbf{Input:} $ L_{A} $, $ L_{B} $\\
	\textbf{Output:} $ \mathcal{C}_{ij} $, $ R_{ij} $\\
	Obtain the positions of transmitter $ i $ and receiver $ j $ of flow $ f\left( i,j \right) $ in the current slot;\\
	Get four boundary abscissa values $ {{x}_{{{P}_{1}}}} $, $ {{x}_{{{P}_{2}}}} $, $ {{x}_{{{Q}_{1}}}} $, and $ {{x}_{{{Q}_{2}}}} $;\\
	Find four corresponding alternate relay nodes to the right around the boundary abscissa value $ P_{1} $, $ P_{2} $, $ Q_{1} $, and $ Q_{2} $;\\
	Let flow $ f\left( i,j \right) $ test the above four nodes, select the node with the fastest transmission rate as the relay node $ \mathcal{C}_{ij} $ in this frame, and assign its rate to $ R_{ij} $;
\end{algorithm}
		
First, input the initial positions $ L_{A} $ and $ L_{B} $ of train A and B. Since flow $ f\left( i,j \right) $ is blocked by obstacles, we start with transmitter $ i $ and receiver $ j $ respectively. If node $ i\left( {{x}_{i}},{{y}_{i}} \right) $ is located in the area directly above the obstacle, that is, $ \left( n-1 \right)len\le {{x}_{i}}\le \left( n-1 \right)len+Bd $, the effective selection range of relay node is the blue area in the current time slot, which can ensure that the flow assisted by relay node does not touch the obstacles. The two left boundaries of the blue area are: the ray formed by node $ i $ and point 4 of the ($ n-1 $)-th obstacle, and the ray formed by node $ i $ and point 2 of the $ n $-th obstacle. They intersect train B at $ p_{1} $ and $ p_{2} $, respectively. The two right boundaries of the blue area are: the ray formed by node $ i $ and point 1 of the $ n $-th obstacle, and the ray formed by node $ i $ and point 3 of the ($ n+1 $)-th obstacle. If two trains drive to the right at different speeds, the right boundary of the blue area will definitely encounter the obstacle in the next time slot. Therefore, the right boundary is invalid, the left boundary within the effective range should be selected as the effective relay node. At this time, no matter the two trains are running at any speed, it can be ensured that they are not blocked by obstacles for a short period of time. It is helpful to achieve effective communication and can ensure the maximum effective transmission time. Then according to the two-point formula in the mathematical method, the abscissa $ {{x}_{{{p}_{1}}}} $ of the left boundary of the effective area on the left side of node $ i $ can be expressed as
\begin{equation}
	{{x}_{{{p}_{1}}}}={{x}_{i}}-\frac{{h}\left( {{x}_{i}}-{{x}_{{4}}} \right)}{{h}-{{y}_{{4}}}},
\end{equation}
where $ {{x}_{{4}}}=\left( n-1 \right)len-nonBd $. The abscissa $ {{x}_{{{p}_{2}}}} $ of the left boundary of its right side effective area can be expressed as
\begin{equation}
	{{x}_{{{p}_{2}}}}=\frac{{h}\left( {{x}_{{2}}}-{{x}_{i}} \right)}{{h}-{{y}_{{2}}}}+{{x}_{i}},
\end{equation}
where $ {{x}_{{2}}}=\left( n-1 \right)len+Bd $.
		
If node $ j $ is within the obstacle blockage range, that is, $ \left( n-1 \right)len\le {{x}_{j}}\le \left( n-1 \right)len+Bd $, it is the same as above. The effective selection range of relay nodes is the green area. We need to select the left boundary of the two effective areas as the effective relay node to maximize the system performance. The two left boundaries of the green area are: the ray formed by node $ j $ and point 2 of the ($ n-1 $)-th obstacle, and the ray formed by node $ j $ and point 4 of the $ n $-th obstacle. They intersect train B at $ q_{1} $ and $ q_{2} $, respectively. Then the abscissa $ {{x}_{{{q}_{1}}}} $ of the left boundary of its left side effective area can be expressed as
\begin{equation}
	{{x}_{{{q}_{1}}}}={{x}_{j}}-\frac{h\left( {{x}_{j}}-{{x}_{{2}}} \right)}{{{y}_{{2}}}},
\end{equation}
where $ {{x}_{{2}}}=\left( n-1 \right)len-nonBd $. The abscissa $ {{x}_{{{q}_{2}}}} $ of its right side effective area can be expressed as
\begin{equation}
	{{x}_{{{q}_{2}}}}=\frac{h\left( {{x}_{{4}}}-{{x}_{j}} \right)}{{{y}_{{4}}}}+{{x}_{j}},
\end{equation}
where $ {{x}_{{4}}}=\left( n-1 \right)len+Bd $.
		
If the node $ i $ is located in a non-obstacle area, that is, $ \left( n-1 \right)len+Bd\le {{x}_{i}}\le nlen $. The abscissa expressions of the left boundary $ p_{3} $ and $ p_{4} $ of the left and right effective areas are the same as those of $ p_{1} $ and $ p_{2} $. Only $ {{x}_{{4}}}$ and $ {{x}_{{2}}}$ need to be set as $nlen-nonBd $ and $ nlen+Bd $.

If the node $ j $ is located in a non-obstacle area, that is, $ \left( n-1 \right)len+Bd\le {{x}_{j}}\le nlen $. The abscissa expressions of the left boundary $ q_{3} $ and $ q_{4} $ of the left and right effective areas are the same as those of $ q_{1} $ and $ q_{2} $. Only $ {{x}_{{2}}}$ and $ {{x}_{{4}}}$ need to be set as $nlen-nonBd $ and $ nlen+Bd $.
		
Therefore, in different time slots, the positions of the transmitter and the receiver of the flow are different, and they will be at the obstacle shelter or nonshelter. The relay selection algorithm is shown in Algorithm \ref{alg:relay}. It requires the position information of two trains, and obtain the positions of transmitter $ i $ and receiver $ j $ of flow $ f\left( i,j \right) $ to be transmitted. In each case, there are four boundary positions, and abscissas are respectively recorded as $ {{x}_{{{P}_{1}}}} $, $ {{x}_{{{P}_{2}}}} $, $ {{x}_{{{Q}_{1}}}} $, $ {{x}_{{{Q}_{2}}}} $. We need to combine the boundary location with the actual scenario to find a suitable relay node to reduce the link outage probability. If the boundary position does not match the position of the existing node, the node looking to the left will definitely encounter obstacles and the link will be blocked. Therefore, we can only look to the right until we encounter existing MR nodes, which will greatly reduce the probability that the relay flow is still blocked. We will get four alternative relay nodes $ P_{1} $, $ P_{2} $, $ Q_{1} $ and $ Q_{2} $. Then make the flow test the four nodes separately, and find the node that makes the flow transmission rate the fastest. We take it as the relay assisted node of the flow in this frame, and then output the relay node $ \mathcal{C}_{ij} $ and the transmission rate $ R_{ij} $ of the flow assisted by $ \mathcal{C}_{ij} $.
		
In this algorithm, in the scheduling phase of each frame, it is necessary to select relay nodes for each blocked flow. In the worst case, we think that all flows are blocked. So in each frame, the computational complexity is $ O\left( F \right) $.
		
\subsection{Transmission Mode Selection Algorithm}

\begin{algorithm}[t]
	\DontPrintSemicolon
	\caption{Transmission mode selection algorithm}\label{alg:mode}
	\textbf{Input:} $ q_{ij} $ for each flow, $ v_{A} $, $ v_{B} $, $ L_{A} $, $ L_{B} $\\
	\textbf{Output:} $ total $\\
	\textbf{Initialization:} $ total=0 $, $ F^{A} $  = $ \varnothing $, $ F^{B} $  = $ \varnothing $\\
	According to $ v_{a} $, $ v_{b} $, $ L_{a} $, $ L_{b} $, calculate the time $ T $ that two trains can maintain communication;\\
	\For{frame t ($ 1\le t\le T $)}
	{
		$ \tau =t*{{T}_{S}}+\left( t-1 \right)*K $;\\
		\For{each flow $ f_{ij} $}
		{
			\If{$ q_{ij} >0 $}
			{
				\If{flow $ f_{ij} $ is blocked at timeslot $ \tau $ and $ \tau +K $}
				{
					Execute Algorithm \ref{alg:relay} for this flow to obtain the relay node $ \mathcal{C}_{ij} $ in this frame and the link transmission rate $ R_{ij} $ under the assistance of $ \mathcal{C}_{ij} $;\\
					\If{$ \mathcal{C}_{ij} \ne i,j $ and $ R_{ij}>0 $}
					{
						$ {{F}^{B}}={{F}^{B}}\cup \left\{ {{f}_{ij}} \right\} $;
					}
				}
				\ElseIf{$ f_{ij} $ is not blocked at slot $ \tau $}
				{
					Calculate the direct link transmission rate $ R_{ij} $ of $ f_{ij} $;\\
					$ {{F}^{A}}={{F}^{A}}\cup \left\{ {{f}_{ij}} \right\} $; $ \mathcal{C}_{ij}=0 $;
				}
			}
		}	
		Start executing the Algorithm \ref{alg:trans};\\
		$ total=total+complete $;\\		
		$ F^{A} $  = $ \varnothing $; $ F^{B} $  = $ \varnothing $.
	}
\end{algorithm}
		
Before starting Algorithm \ref{alg:mode}, we should first obtain the traffic of all flows to be transmitted. Then, according to the speeds $ v_{A} $, $ v_{B} $ and initial positions $ L_{A} $, $ L_{B} $ of the two trains driving in the same direction, calculate the time $ T $ for two trains to remain in the communication range.
		
In line 5, we begin to select the transmission mode for all flows in each frame. Set $ \tau $ as the first time slot of the current frame transmission phase. Lines 6-14 is describe process of selecting the transmission mode for each flow with the traffic to be transmitted. Here, two empty sets $ F^{A} $ and $ F^{B} $ are defined in advance, which are the sets of direct mode flows and relay mode flows to be transmitted. If a flow with remaining traffic is blocked at both the first and last timeslot in the transmission phase, it means that it is always blocked with a high probability in the current frame, so relay assistance is required. In line 10, the above relay selection algorithm is execated for this flow to obtain the optimal relay assisted node $ \mathcal{C}_{ij} $ in this frame and the relay-assisted transmission rate $ R_{ij} $. In line 11, if relay node $ \mathcal{C}_{ij} $ is different from the transceiver  of the flow itself, and $ R_{ij} $ is greater than 0, this is an effective relay node. This flow is put into set $ F^{B} $, which indicates that this flow has the opportunity to be transmitted. For relay mode flows that do not meet the conditions, we will give up scheduling them.
		
In line 12, if the flow is not blocked at the first time slot of the transmission phase, the direct mode can be used for transmission for a period of time. Because the link situation is dynamic and uncertain. At the same time, in order to reduce the occupation of node states and maximize the system performance, we believe that the direct transmission mode should be selected as far as possible for the flow that can use the direct transmission. Then the flow is put into set $ F^{A} $ and wait for transmission. At the same time, set $ \mathcal{C}_{ij} $ to 0, indicating that there is no relay node in flow $ f_{ij} $. In line 15, the transmission mode has been allocated for all flows. Next, start the scheduling algorithm in each frame, that is the Algorithm \ref{alg:trans}.

In lines 17-18, the number of completed flows of each frame is accumulated, and $ F^{A} $ and $ F^{B} $ are set to empty to avoid affecting the scheduling of other frames. When $ T $ cycles are completed, the final result $ total $ is the maximum number of communication flows that can be completed in the T2T communication system.
		
This algorithm is to select the appropriate transmission mode for each flow in the frame, and it needs to judge $ F $ flows. The computational complexity in each frame is $ O\left( F \right) $. Algorithm \ref{alg:mode} is parallel to the Algorithm \ref{alg:relay}, so the complexity of these two algorithms is still $ O\left( F \right) $.

\subsection{Transmission Scheduling Algorithm}
First, we need to obtain the direct link set $ F^{A} $ and relay link set $ F^{B} $ that can be transmitted in the current frame obtained by Algorithm 2. Calculate the number of time slots $ {{t}_{ij}} $ for the flow by its traffic $ q_{ij} $ and transmission rate $ R_{ij} $, which can be expressed as
\begin{equation}
	{{t}_{ij}}=\frac{{{q}_{ij}}}{{{R}_{ij}}*\Delta t}.
\end{equation}

\begin{algorithm}[t]
	\DontPrintSemicolon
	\caption{Transmission scheduling algorithm}\label{alg:trans}
	\textbf{Input:} $ q_{ij} $, $ R_{ij} $ for each flow, $ F^{A} $, $ F^{B} $\\
	\textbf{Output:} $ complete $\\
	\textbf{Initialization:} $ complete=0 $, $ F^{re} $  = $ \varnothing $, $ TRS $  = $ \varnothing $, $ S^{t}_{i} $ = $ S^{r}_{i} $ = 0 for each node $ i $.\\
	Arrange the flows in $ F^{A} $ and $ F^{B} $ in ascending order of time slots;\\
	$ {{F}^{re}}={{F}^{A}}\bigcup {{F}^{B}} $;\\
	$ choose=1 $;\\
	\For{slot k ($ 1\le k\le K $)}
	{
		\If{$ choose=1 $}
		{
			\For{each flow $ f_{ij} $ $\in$ $ F^{re} $}
			{
				\If{$ \mathcal{C}_{ij} =0 $ and $ S^{t}_{i} $ = $ S^{r}_{j} =0 $  }
				{
					$ TRS=TRS\cup \left\{ {{f}_{ij}} \right\} $;\\
					$ S^{t}_{i} $ = $ S^{r}_{j} =1 $;
				}
				\ElseIf{$ \mathcal{C}_{ij} \ne 0 $ and $ S^{t}_{i} $=$ S^{r}_{\mathcal{C}_{ij}} $=$ S^{t}_{\mathcal{C}_{ij}} $=$ S^{r}_{j}=0 $  }
				{
					$ TRS=TRS\cup \left\{ {{f}_{ij}} \right\} $;\\
					$ S^{t}_{i} $=$ S^{r}_{\mathcal{C}_{ij}} $=$ S^{t}_{\mathcal{C}_{ij}} $=$ S^{r}_{j}=1 $;
				}
			}
			$ choose=0 $;
		}
		\For{each flow $ f_{ij} \in $ TRS}
		{
			\If{$ R_{ij}>0 $}
			{
				Calculate the remaining traffic $ q_{ij} $ of this flow;\\
				\If{$ q_{ij}<0 $}
				{
					$ complete=complete+1 $;\\
					\textbf{Go to line 24}
				}
			}
			\ElseIf{$ R_{ij}=0 $}
			{
				Delete flow $ f_{ij} $ in sets $ TRS $ and $ F^{re} $, and set the corresponding transceiver status of the flow to 0;\\
			}
		}
	}
\end{algorithm}		
In line 4, we arrange the flows in sets $ F^{A} $ and $ F^{B} $ in ascending order of the number of required time slots. Combine the two sets into $ F^{re} $ in the order of $ F^{A} $ first and then $ F^{B} $. In this way, the direct transmission flows will be scheduled first, and each node's resource will be fully utilized to complete more flow transmission. In line 6, we set a flag variable $ choose $, which indicates whether to add new flows to the set $ TRS $. In lines 7-24, the algorithm aims to complete the transmission of as many flows as possible in the current frame. When the flag variable $ choose $ is 1, the utilization rate of the node's resource is not enough in the current slot, and more flows can be added in $ TRS $. The rules for adding flows are shown in lines 9-15. We select flows from set $ F^{re} $ and add them to set $ TRS $ in ascending order of time slots. For any flow $ f_{ij} $ in $ F^{re} $, if it uses the direct mode, and the transmitting state of node $ i $ and the receiving state of node $ j $ are idle, then this flow can be put into the $ TRS $ set and wait to be transmitted. At the same time, set $ S^{t}_{i} $ and $ S^{r}_{j} $ to 1, which indicates that the node resources have been occupied. In the current slot, other flows can no longer use the transmitting end of $ i $ and the receiving end of $ j $. If it is transmitted by relay, similarly, we judge whether the transmitting state of node $ i $, the transmitting state and receiving state of node $ \mathcal{C}_{ij} $, and the receiving state of node $ j $ are idle. If the conditions are met, put the flow into set $ TRS $, and set the corresponding node states to 1. After judging all the flows in $ F^{re} $, we set the flag variable to 0, and then start the following actual transmission.
		
For each flow in $ TRS $ in the current timeslot, if the actual transmission rate is greater than 0, we will calculate the remaining traffic after transmission. In line 20, when the remaining traffic is less than 0, it means that this flow has just been completely transmitted. Then, we add a flow that has completed transmission in the current frame. In line 24, for the flow that has just completed its transmission and the flow whose transmission rate is 0 in the current timeslot, we should delete it from the set $ TRS $ and $ F ^ {re} $. Because the flow that has completed transmission does not need to be transmitted again. The flow whose rate is equal to 0 indicates that the current timeslot is blocked, then most of the next timeslots will still be blocked. In order to make full use of all node states, we decide to give up the transmission of this flow in the current frame. Then set the transmitting and receiving status of the corresponding node to 0 to provide transmission opportunities for other flows.
		
This algorithm aims to achieve the transmission scheduling of all flows in the current frame and maximize the number of completed flows. First, we need to arrange the flows in the set in ascending order according to the number of time slots required. The computational complexity is $ O\left( F{{\log }_{2}}F \right) $. Then, in each time slot of the actual transmission phase, it is necessary to select flows from set $ F^{re} $ and add them to set $ TRS $ for transmission in turn. The two parts are parallel $ for $ loops, so the operation complexity is $ O\left( FK \right) $. To sum up, we assume that there are $ T $ communication frames, and the computational complexity of the heuristic algorithm can be expressed as $ O\left( TF\left( {{\log }_{2}}F+K \right) \right) $. For optimal solution found by an exhaustive search, the computational complexity is $ O\left( {{\left( 2N-1 \right)}^{FTK}} \right) $, which does not work for pratical systems.
		
\section{Performance Evalution}\label{S6}
In this section, in order to evaluate the performance  of the proposed heuristic algorithm with respect to the number of completed flows and system throughput, we compare it with three baseline schemes in simulation.
\subsection{Simulation Setup}
		
In the simulation, there are two trains running in the same direction but at different speeds. The train length $ L_{t} $ is 200 meters, with 8 carriages, and each train has 16 MRs evenly distributed on the roof. Trains run on two different tracks. For ease of explanation, the positions of the two trains can be represented by three-dimensional coordinates $ \left( x,y,z \right) $. The position coordinates of the two locomotives are represented as $ \left( {{x}_{A}},{{y}_{A}},{{z}_{A}} \right) $ and $ \left( {{x}_{B}},{{y}_{B}},{{z}_{B}} \right) $, respectively. Here, the communication distance threshold $ dis $ is defined as the longest distance that two trains can maintain communication with each other, and it is necessary to achieve communication between all MRs on these two trains. That is to say, the MR located at the forefront can communicate with the MR located at the end. Therefore, when the distance between the two locomotives $ \left| {{x}_{A}}-{{x}_{B}} \right|=dis-{{L}_{t}} $, it indicates that the two trains  have reached the communication threshold. $ \left| {{y}_{A}}-{{y}_{B}} \right| $ is the distance between two tracks, which we set to 150 meters in the simulation. Due to the fact that all trains are located at the same height, $ \left| {{z}_{A}}-{{z}_{B}} \right| $. For the T2T communication system, we only consider communication flows between MRs of different trains, that is, the transmitter and receiver of each flow must be located on two different trains. The traffic demand to be transmitted for each flow is randomly generated between 30-50Mb. The communication time of the system is determined by the initial position and speed of the two trains (It is assumed that the speed of the trains remain constant). According to the parameters in \cite{Wang}, we set the scheduling phase time to 850 $ \mu $s and the number of transmission time slots to 2000 in each frame in the simulation. Since all MRs use FD communications, there will be self-interference. The higher the SI cancellation level, the smaller the $\beta$ parameter, and the easier it is for the flow to meet the transmission requirements. Here, we set the SI cancellation parameter $\beta$ to -130 dB. Other parameters are shown in Table I.
		
The system adopts the real directional antenna model in the IEEE 802.15.3c standard, with a Gaussian-shaped main lobe and constant-gain side lobe. The antenna gain $ G(\theta $) is expressed as
		\begin{equation}
			G(\theta) =
			\begin{cases}
				G_0-3.01\times\left(\frac{2\theta}{\theta_{\mbox{\tiny{-3dB}}}}\right)^2, &\mbox{$0^{\circ}\le\theta\le\theta_{ml}/2$},\\
				G_{sl}, &\mbox{$\theta_{ml}/2<\theta\le180^{\circ}$},
			\end{cases}
		\end{equation}
where $ {\theta} $ takes a value in  $[0^\circ,180^\circ]$. $ {\theta_{\mbox{\tiny{-3dB}}}} $ is the half-power beamwidth. The main lobe width $ \theta_{ml} $ can be expressed as $\theta_{ml}=2.6\times\theta_{\mbox{\scriptsize{-3dB}}}$. $G_0$ is the maximum antenna gain, which can be expressed as
		\begin{equation}
			G_0=10\mbox{log}(1.6162/\mbox{sin}(\theta_{{\mbox{\scriptsize{-3dB}}}}/2))^2.
		\end{equation}
		$ G_{sl} $ is the side lobe gain, which can be expressed as
		\begin{equation}
			G_{sl}=-0.4111\times\mbox{ln}(\theta_{\mbox{\scriptsize{-3dB}}})-10.579.
		\end{equation}
		
		\begin{table}[t]
			\caption{Simulation parameters} \label{table:parameter setting}
			\centering  
			\begin{tabular}{lccc}
				\hline
				\textbf{Parameter}  &\textbf{Symbol}&\textbf{Value}\\
				\hline
				Transmitting power &$ {P}_{t} $&1000mW\\
				Transmission frequency &$ f $&28GHz\\
				Bandwidth&$ W $&1200MHz\\
				Background noise&$ {N}_{0} $&-134dBm/MHz\\
				Path loss index&$ n $&2\\
				Half power beam width&$ \theta_{\mbox{\scriptsize{-3dB}}} $&$ 30^\circ $\\
				Slot time&$ \Delta t $&$ 18\mu s $\\
				Scheduling phase time&$ {T}_{s} $&$ 850\mu s $\\
				SI cancellation parameter& $\beta$ & ${10^{-13}}$\\
				\hline
			\end{tabular}\\
		\end{table}
		
This paper will measure the system performance of the number of completed flows and system throughput. To verify the performance advantages of the proposed algorithm, we compare the heuristic algorithm with the following three baseline schemes.\\
1) \emph{\textbf{Direct transmission}}: All flows are transmitted in the direct mode and are scheduled in ascending order of the number of time slots required for each flow to complete its transmission during the transmission phase. Transmission of a blocked flow in the current frame will be aborted.\\
2) \emph{\textbf{Hybrid selective transmission}}: It combines the direct transmission mode and the relay transmission mode, selects the transmission mode with the highest transmission rate within the current frame for each flow during the frame scheduling phase. During the transmission phase, the transmission scheduling of this scheme is in ascending order of the number of time slots required for each flow to complete its transmission.\\
3) \emph{\textbf{Random transmission}}: It combines the direct transmission mode and the relay transmission mode, randomly selects one of the transmission modes for all flows during the scheduling phase of this frame. If the relay mode is selected, it is also necessary to randomly select a node as the relay assistance for the flow. During the transmission phase of the frame, priority is given to flows transmitted using direct mode, followed by flows transmitted using relay mode.
		
\subsection{Comparison With Existing Schemes}
1) \emph{Communication threshold}
\begin{figure}[t]
	\begin{minipage}[t]{1\linewidth}
		\centering
		\includegraphics[width=1\columnwidth]{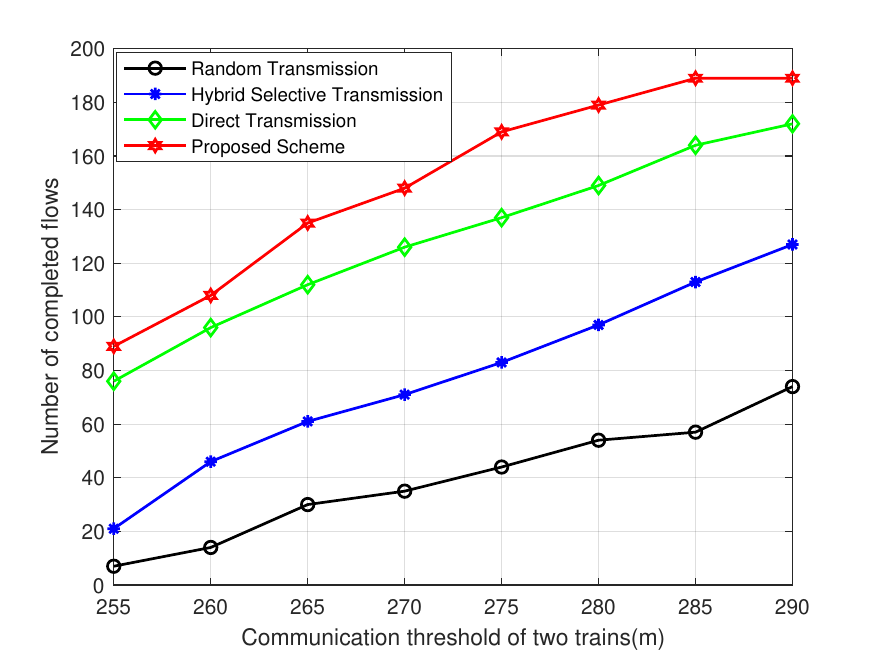}
	\end{minipage}%
	\caption{The influence of communication threshold on the number of completed flows}
	\label{fig:threshold1_total} 
	\vspace*{-3mm}
\end{figure}

\begin{figure}[t]
	\begin{minipage}[t]{1\linewidth}
		\centering
		\includegraphics[width=1\columnwidth]{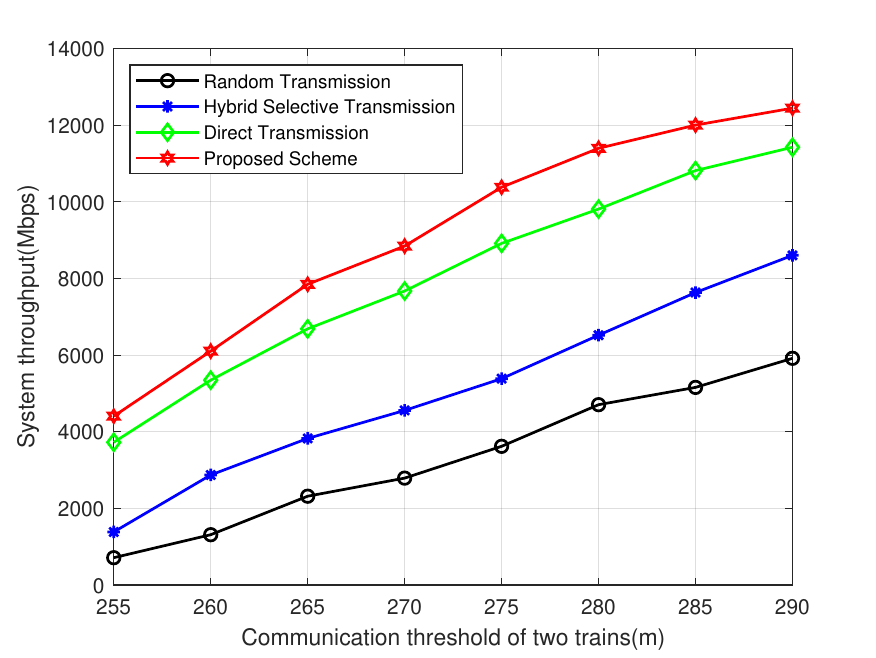}
	\end{minipage}%
	\caption{The influence of communication threshold on system throughput}
	\label{fig:threshold1_thput} 
	\vspace*{-3mm}
\end{figure}
In the simulation, we set the two trains to have the same initial position, train A has a speed of 300km/h, and train B has a speed of 150km/h. The blockage probability is 40\%. There are 200 flows randomly distributed at 30-50Mb to be transmitted. The communication threshold determines the time when the two trains can maintain communication. The communication time increases with the increase of the communication threshold, which means that more flows can be transmitted. It can be seen from Figs. \ref{fig:threshold1_total} and \ref{fig:threshold1_thput} that in these four transmission modes, the number of completed flows and system throughput are gradually increasing with the increase of communication threshold. Next, we compare the performance of schemes. Obviously, the performance of the proposed scheme is higher than those of these baseline schemes, since it can complete more flows and achieve greater throughput. Compared with each baseline scheme, the number of completed flows has increased by 17\%, 124\% and 427\%, respectively. The overall throughput is increased by 15\%, 102\% and 244\%, respectively. The above results show that it is not advisable to choose the fastest rate transmission mode for each flow in the current frame, because the overall performance of the hybrid selective transmission mode is usually significantly lower than that of the direct transmission mode. This is because channel conditions change dynamically, and some blocked links may be recovered in subsequent frames. If the relay mode must be used for transmission in the current frame, it will occupy too much transceivers, hinder the transmission of other direct flows, and affect the overall performance. Therefore, we cannot just pursue the relay mode, but need to combine the direct mode with the relay mode, which is the essence of the proposed algorithm. When the communication threshold is not large, the performance of the heuristic algorithm is 20\% higher compared with the best baseline scheme. In the case of high threshold, the performance gain is about 15\%. This is because when the communication threshold is very high, all schemes can maintain communication for a long time. Even if there is serious path loss, all flows still have enough time to complete the transmission, which is not a good scenario to take advantage of the heuristic algorithm. When the threshold is lower, the heuristic algorithm can comprehensively allocate the limited resources, make full use of the node receiving and transmitting states, and combine the traffic demand of each flow and dynamic channel conditions to maximize the number of transmission flows and system throughput.
		
2) \emph{Relative speed of two trains}
\begin{figure}[t]
	\begin{minipage}[t]{1\linewidth}
		\centering
		\includegraphics[width=1\columnwidth]{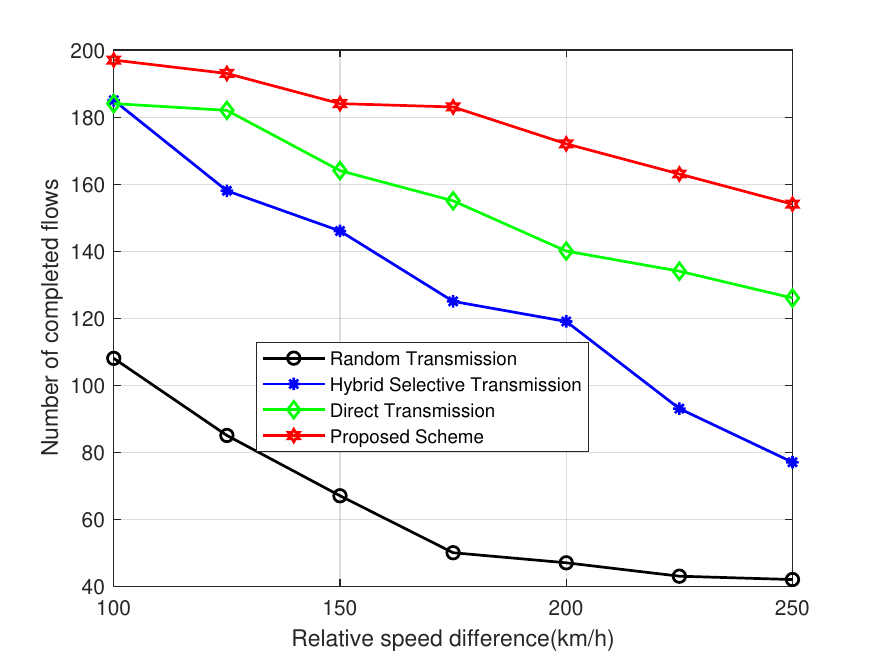}
	\end{minipage}%
	\caption{The influence of relative speed difference on the number of completed flows}
	\label{fig:relaspeed3_total} 
	\vspace*{-3mm}
\end{figure}

\begin{figure}[t]
	\begin{minipage}[t]{1\linewidth}
		\centering
		\includegraphics[width=1\columnwidth]{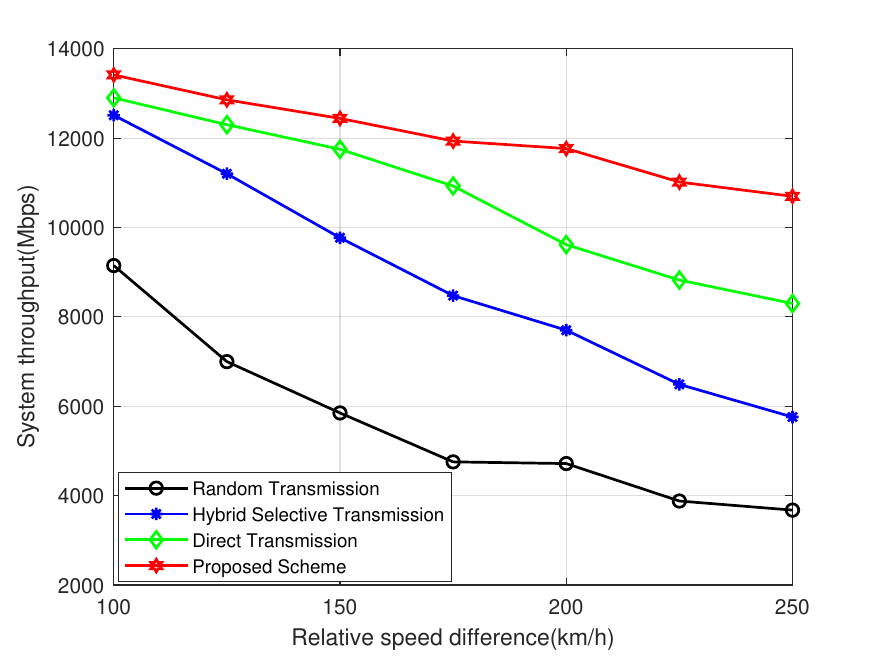}
	\end{minipage}%
	\caption{The influence of relative speed difference on system throughput}
	\label{fig:relaspeed3_thput} 
	\vspace*{-3mm}
\end{figure}
In the simulation, we set the two trains to have the same initial position. The article only considers the communication between two trains running in the same direction. We set the speed of train A to 300km/h, the speed of train B is reduced from 200km/h to 50km/h uniformly, and the simulation test is conducted at every 25km/h. The blockage probability in this scenario is 40\%. There are 200 flows randomly distributed at 30-50Mb to be transmitted. The horizontal communication threshold is 250m. The greater the relative speed difference, the less time for maintaining communication. As shown in Figs. \ref{fig:relaspeed3_total} and \ref{fig:relaspeed3_thput}, when the relative speed difference gradually increases from 100km/h to 250km/h, the number of completed flows shows a downward trend as a whole. Compared with the four schemes, the performance of the heuristic algorithm is still significantly higher than those of the baseline schemes. The number of completed flows is increased by 15.7\%, 45\%, and 204\%, respectively compared with the baseline schemes. The system throughput is increased by 11.6\%, 27.4\%, and 54.2\%, respectively. We find that when the relative speed difference is low, the performance gain of the heuristic algorithm is not obvious, which is the same as the previous study. Sufficient resources enable each scheme to have more time to transmit flows. In the case of tight communication time, the proposed algorithm can better realize resource allocation, maximize node status utilization, and achieve a performance gain by 22\%, compared with the best baseline scheme.
		
3) \emph{Actual speed of two trains}
\begin{figure}[t]
	\begin{minipage}[t]{1\linewidth}
		\centering
		\includegraphics[width=1\columnwidth]{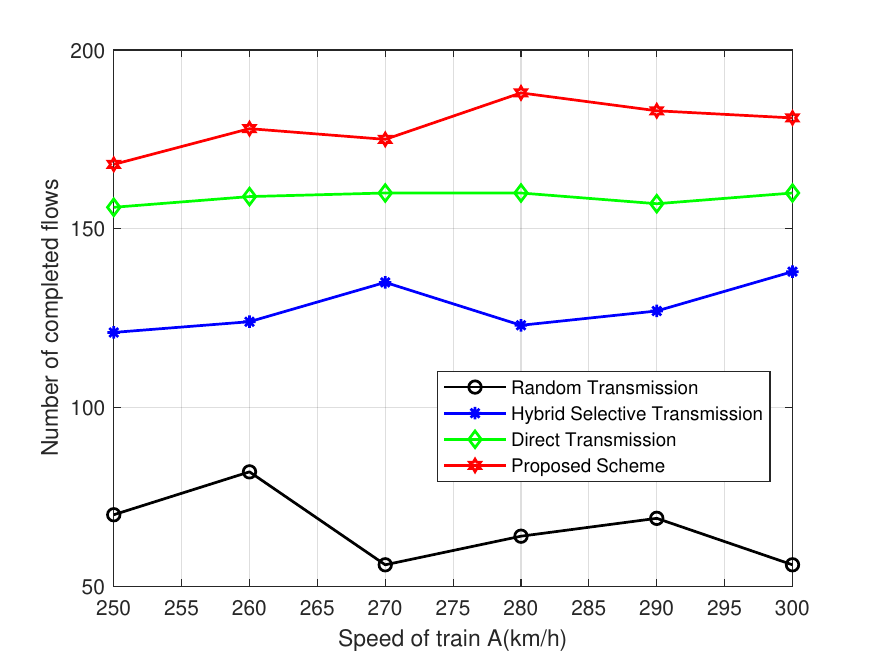}
	\end{minipage}%
	\caption{The influence of actual speed on the number of completed flows}
	\label{fig:Aspeed1_total} 
	\vspace*{-3mm}
\end{figure}

Next, the influence of the absolute speed of two trains on the number of completed flows is analyzed. In the simulation, the initial positions of two trains are the same. We make train A faster than train B and set the relative speed difference to 150km/h. We will uniformly increase the speed of train A from 250km/h to 300km/h, and conduct the simulation every 10km/h. The blockage probability is 40\%. There are 200 flows to be transmitted, and the traffic demand is randomly distributed between 30-50Mb. The horizontal communication threshold is 250m. It can be seen that the performance of the proposed algorithm is always better than that of the baseline algorithms, and the completed flow rate is 13\% higher than that of the best baseline scheme. The number of completed flows is positively related to the number of communication frames. While the number of communication frames depends on the initial position and relative speed difference of the two trains, and is not directly related to the actual speed. It can be seen from Fig. \ref{fig:Aspeed1_total} that under the premise of a relative speed difference of 150km/h and a blockage probability of 40\%, when the speed of train A is 280km/h, the number of completed flows is the largest, and the performance is 17.5\% higher than that of the direct mode.
		
4) \emph{Relative initial position of two trains}
\begin{figure}[t]
	\begin{minipage}[t]{1\linewidth}
		\centering
		\includegraphics[width=1\columnwidth]{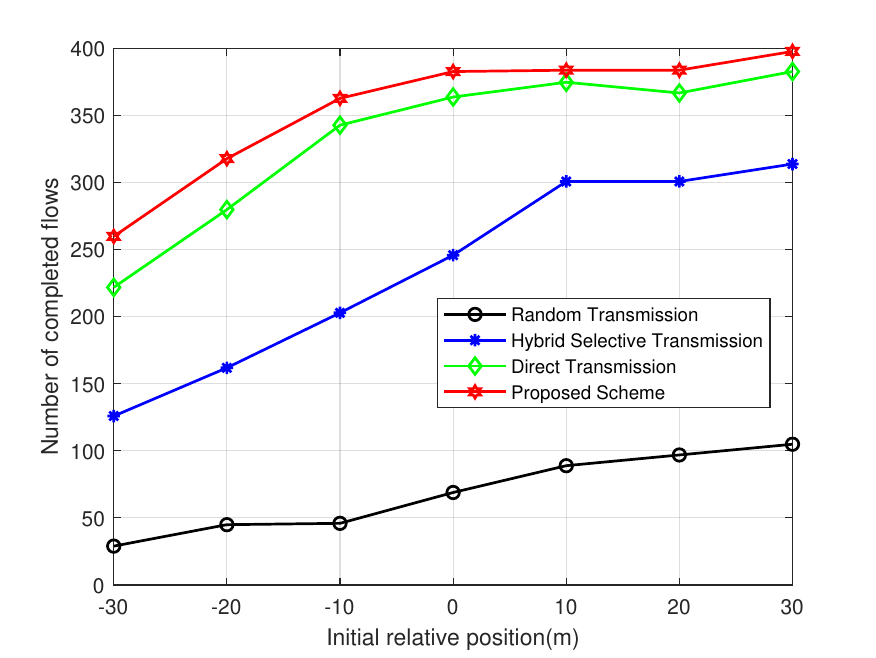}
	\end{minipage}%
	\caption{The influence of initial relative position on the number of completed flows}
	\label{fig:loc6_total} 
	\vspace*{-3mm}
\end{figure}
\begin{figure}[t]
	\begin{minipage}[t]{1\linewidth}
		\centering
		\includegraphics[width=1\columnwidth]{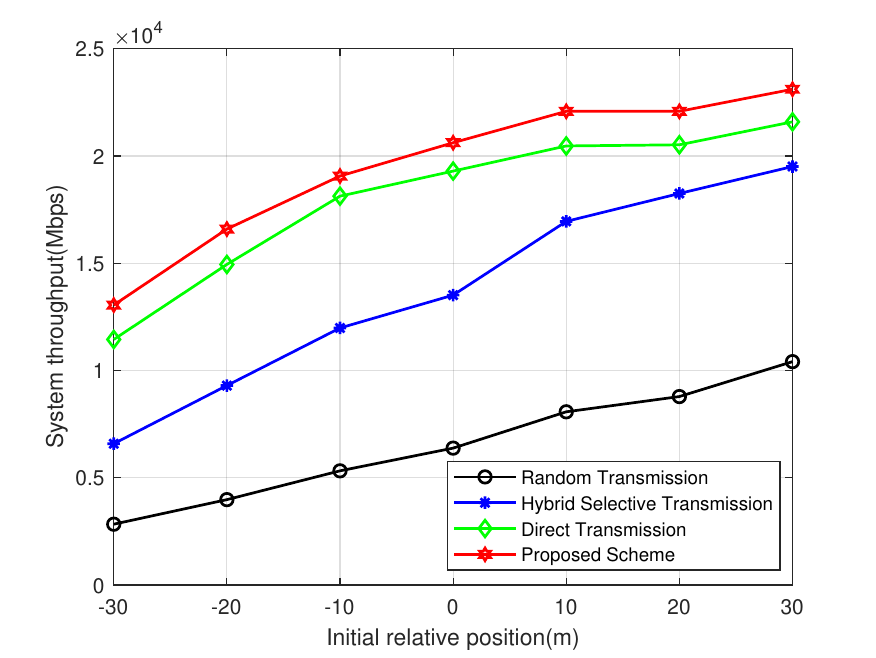}
	\end{minipage}%
	\caption{The influence of initial relative position on system throughput}
	\label{fig:loc6_thput} 
	\vspace*{-3mm}
\end{figure}

The influence of the initial relative position of two trains on the system performance is analyzed. In the simulation, the speed of train A is 300km/h and that of train B is 150km/h. The initial position of train A is at 500m. We vary the initial position of train B. The blockage probability is 40\%. There are 400 flows to be transmitted, and the traffic is randomly distributed between 30-50Mb. The horizontal communication threshold is 250m. As shown in Figs. \ref{fig:loc6_total} and \ref{fig:loc6_thput}, the abscissa is the relative position difference between two trains. The initial position of train B is between 470-530 meters, and the simulation is conducted every 10 meters. With the increase of the initial relative position difference between the two trains, the number of communication frames will gradually increase, and the overall performance will show an upward trend. The availability of MR is improved with the increase of the number of flows to be transmitted. Compared with the initial set of 200 flows to be transmitted, more nodes will participate in the transmission of each timeslot, and the number of idle nodes will be further reduced. Therefore, when the number of flows to be transmitted is different, the number of flows that can be completed within the same communication frames will varies greatly. The number of completed flows of the heuristic algorithm is 8\%, 60\% and 492\% higher, respectively, compared with the three baseline schemes. The system throughput is improved by 8\%, 31.7\% and 67.4\%, respectively. On the premise that train A is faster than train B, if train B is left behind train A at the initial position, the number of communication frames will be smaller. The proposed algorithm can achieve transmission scheduling of more flows in a short time. The effect is obvious, and the completion rate is 17\% higher than that of the direct mode; If the initial position of B is in front of A, it means that train A will catch up train B, so the number of communication frames will become larger. Even a flow with a very slow transmission rate is more likely to complete its transmission. The advantages of the algorithm are not so obvious, and the completion rate is 5\% higher than that of the direct mode.
		
5) \emph{Blockage probability}
\begin{figure}[t]
	\begin{minipage}[t]{1\linewidth}
		\centering
		\includegraphics[width=1\columnwidth]{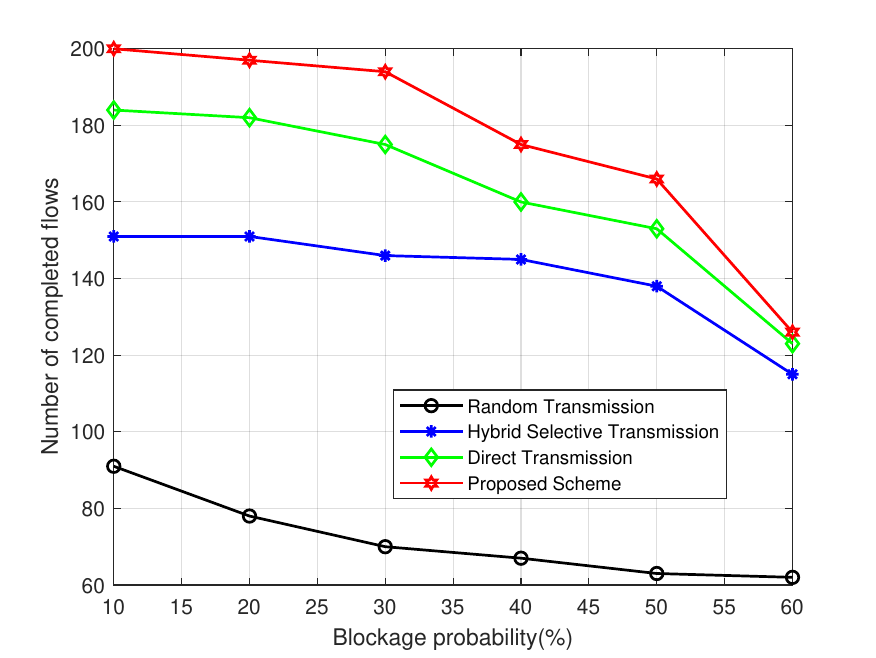}
	\end{minipage}%
	\caption{The influence of obstacle blockage probability on the number of completed flows}
	\label{fig:block3_total} 
	\vspace*{-3mm}
\end{figure}

In the simulation, the two trains have the same initial position, train A has a speed of 300km/h, and train B has a speed of 150km/h. There are 200 flows randomly distributed between 30-50Mb to be transmitted. The horizontal communication threshold is 250m. As shown in Fig. \ref{fig:block3_total}, the number of completed flows decreases gradually with the increase of obstacle blockage probability. Compared with the three baseline algorithms, the heuristic algorithm achieves a gain of 8.2\%, 24.8\%, and 178\%, respectively. If the obstacle blockage probability is too high, i.e. reaching 60\%, it indicates that the channel conditions are too poor. At this time, the link using relay mode will also be easily blocked, making the algorithm not much useful. The performance is only improved by 3\%. When the blockage probability is within a reasonable range, the proposed algorithm will greatly improve the system performance, and the completion rate is 11\% higher than that of the direct mode.

\section{Conclusion}\label{S7}
In this paper, we studied the robust transmission scheduling problem in mmWave T2T communication systems under blockage. In this paper, MRs on top of the train were used to overcome the link blockage problem, thus improving the NLOS link to LOS links, and enhancing the system robustness. We designed a low complexity heuristic algorithm for T2T communication system, which consists three components: relay selection, transmission mode selection, and transmission scheduling. This heuristic algorithm can not only greatly reduce the computational complexity, but also combine the channel dynamic changes and traffic requests well, so that each flow can be transmitted in the right mode at the right time to maximize the system performance. Finally, we compared the proposed algorithm with three baseline schemes and analyzed the system performance under different system parameters. The results showed that the proposed algorithm could make good use of the node's full-duplex state and time resources within a reasonable communication threshold. It compromises system performance and computational complexity to achieve more completed flows and system throughput. Compared with three baseline scheme, it can have a performance gain of 22\%.

\end{document}